\documentclass[useAMS,usenatbib,letter]{mn2e_v2}
\usepackage{psfig} 
\usepackage[dvips]{graphicx}

\def\xmm{{\sl XMM-Newton}}
\def\asca{{\sl ASCA}}
\def\rxte{{\sl RXTE}}
\def\ark{{Ark~564}}

\title[Time Lags in Ark 564]{Spectral-timing evidence for a Very High State in the Narrow Line Seyfert 1 Ark 564}  
\author[P. Ar\'evalo, I. E. Papadakis, P. Uttley, I. M. McHardy, W. Brinkmann]{P. Ar\'evalo$^{1}$\thanks{E-mail: patricia@astro.soton.ac.uk}, I. E. Papadakis$^{2}$ P. Uttley$^{3}$, I. M. McHardy$^{1}$, W. Brinkmann$^{4}$ \\ 
$^1$School of Physics and Astronomy, University of Southampton, Southampton S017 1BJ, UK\\
$^2$Physics Department, University of Crete, P.O.Box 2208, 71003 Heraklion, Greece\\
$^3$Astronomical Institute `Anton Pannekoek', University of Amsterdam,
Kruislaan 403, 1098 SJ, Amsterdam, the Netherlands\\
$^4$Max-Planck-Institut f\"ur extraterrestrische Physik, Postfach 1312, D-85741 Garching, Germany\\ 
}

\begin{document}
\date{Received /Accepted}
\pagerange{\pageref{firstpage}--\pageref{lastpage}} \pubyear{2006}

\maketitle
\label{firstpage}

\begin{abstract}
  We use a 100 ks long \xmm\ observation of the Narrow-Line Seyfert 1
  galaxy \ark\ and combine it with the month-long monitoring of the
  same source produced by \asca , to calculate the { phase} lags
  and coherence between different energy bands, over frequencies $\sim
  10^{-6}$ to $10^{-3}$ Hz. This is the widest frequency range for
  which these spectral-timing properties have been calculated
  accurately for any AGN.  The 0.7--2 and 2--10 keV \asca\ light
  curves, and the \xmm\ light curves in corresponding energy bands,
  are highly coherent ($\sim 0.9$) over most of the frequency range
  studied. We observe time lags between the energy bands, increasing
  both with time-scale and with energy separation of the bands. The
  time lag spectrum shows a broad peak in the
  $10^{-5}-5\times 10^{-4}$ Hz frequency range, where the time lags
  follow a power law slope $\sim -0.7$.  Above $\sim 5\times 10^{-4}$
  Hz the lags drop below this relation significantly. This change in
  slope resembles the shape of the lag spectra of black hole X-ray
  binaries (BHXRB) in the very high or intermediate state. The lags
  increase linearly with the logarithm of the separation of the energy
  bands, which poses one more similarity between this AGN and BHXRBs.
\end{abstract}


\section{Introduction}

There is growing evidence that Active Galactic Nuclei (AGN) behave
like scaled-up versions of black hole X-ray binaries (BHXRBs), because
of the similar X-ray variability characteristics and spectral-scaling
properties in both types of system (e.g. \citealt{Merloni,3227}).  It
is therefore tempting to assume that AGN should be found in various 
accretion states, similar to the BHXRB high/soft and low/hard states
and possibly also the transitional very high or intermediate states.
If the longest variability time-scales in these objects follow the
linear scaling with black hole mass seen on shorter time-scales
(e.g. \citealt{McHardyMCG}), then the state transitions seen in BHXRBs
on time-scales of hours and longer would occur in AGN on time-scales
of thousands to millions of years. Therefore, we might expect to see
different states in different AGN, but changes in state in a single
AGN are unlikely to be seen within a human lifetime. 

In BHXRBs, the different states can be identified by their distinct
spectral and timing properties (e.g. \citealt{McClintock}). By
measuring the X-ray variability power spectral density (PSD),
\citet{McHardy4051}, \citet{McHardyMCG} and \citet{3227} have shown
that the Seyfert galaxies NGC~4051, MCG--6-30-15 and NGC 3227 resemble
BHXRBs in the high/soft state, while other AGN, such as NGC~3783
\citep{Markowitzark} and NGC~4258 \citep{4258} may correspond to
BHXRBs in the low/hard state. It has been argued that the Narrow Line
Seyfert~1 \ark\ resembles BHXRBs in the rarer very high state due to
its very high accretion rate \citep{Papadakis_ark} and its
doubly-broken PSD shape, where the separation of the breaks is too
broad to reconcile this PSD with that of the BHXRB Cyg~X-1 in the
low/hard state \citep{Papadakis_ark,doneger}. In this paper, we will
use the cross spectrum (i.e. time lags and coherence between different
energy bands) to compare this AGN with BHXRBs in different states.

 \ark\ is a nearby, X--ray bright, Narrow Line Seyfert 1 galaxy
(NLS1). These objects constitute a subclass of active galactic nuclei
(AGN) that can exhibit very rapid and large amplitude X-ray variability
\citep[e.g.][]{Boller}. In the summer of 2001 \ark\ was observed
continuously by the \asca\ X-ray observatory for more than a
month. This observation was part of a broad-band reverberation mapping
campaign \citep[]{Turner, Shemmer01}. Using the contemporaneous data
obtained during the multi-wavelength campaign in 2001, \citet{Romano}
constructed the broad-band spectral energy distribution of \ark. They
found a bolometric luminosity of $\sim 10^{45}$ ergs s$^{-1}$, which,
for a black hole (BH) of $\sim 10^{7}$ solar masses implies an
accretion rate close to the Eddington limit.

\citet{Pounds} and \citet{Papadakis_ark} have studied the X--ray flux
variability of \ark\ using two-year long, {\it Rossi X-ray Timing
Explorer} (\rxte) monitoring data and the 1 month long \asca\ light
curves. \citet{Pounds} detected a break in the PSD at a frequency
$\sim 8.7\times 10^{-7}$ Hz, which was later recaculated by
\citet{Markowitzark}, who corrected for aliasing and red-noise leak
effects and found the break to be at a frequency $\sim 1.5\times
10^{-6}$. On the other hand, \citet{Papadakis_ark} detected a second
break at $\sim 2\times 10^{-3}$ Hz, which corresponds to a time scale
of $\sim 500$ sec, almost 2000 times smaller than the long time scale
detected by \citet{Pounds}. Although the overall shape of the \ark\
X-ray PSD is similar to that seen in Cyg X-1 in its low/hard state,
the large difference between the two frequency breaks strongly argues
against this possibility.

In this paper, we use a new \xmm\ observation of \ark\, combined with
the month-long observation performed by \asca, to study the time lag
and coherence functions between light curves of various energy bands
and use our results to investigate the X--ray state in which \ark\
might operate. An energy spectral analysis of the \xmm\ observations
will be presented by Papadakis et al. ({\sl A\&A submitted}) while
results from a PSD analysis, using archival, long
\rxte\ and \asca\ light curves, together with the new \xmm\ data, will
be presented by M$^c$Hardy et al. ({\sl in preparation}).

The paper is organised as follows: We briefly describe the data
reduction in Section \ref{data} and calculate the spectral-timing
properties of the light curves in
Section \ref{xs}. In Section \ref{energy}, we investigate the energy
dependence of the time lags. We compare the lag spectra with
other AGN and with BHXRB lag spectra in Section \ref{discussion} and
summarise our conclusions in Section \ref{conclusions}.

\section{The Data}
\label{data}
The 100 ks long, continuous exposure provided by \xmm\ and the
month-long, but periodically interrupted, observation from \asca\
produce data sets that probe complementary time-scale ranges.  In the
present study we combine both data sets to cover the widest possible range.

\subsection{\xmm}

\ark\ was observed by \xmm\ for 100 ks on 2005 January 5 and 6, during
revolution 930. We used data from the European Photon Imaging Cameras
(EPIC) PN and MOS instruments. The PN camera was operated in Small
Window mode, using the medium filter, for a total exposure length of 98.8
ks. Source photons were extracted from a $\sim 2 \arcmin \times 2
\arcmin$ square region and the background was selected from a
source-free region of equal area on the same chip. We selected single
and double events, with {\it quality flag}=0. The source average count rate
in the 0.2--10 keV band is $\sim 28$ c/s. The data showed no indication
of pile-up when tested with the {\it XMM-SAS} task {\it epatplot}. The
background average count rate was $\sim 0.17$ c/s, and stayed practically
constant throughout the exposure.

Both MOS cameras were operated in the Prime Partial Window 2 imaging
mode, using the medium filter, for a total exposure length of 99.1 ks.
Source photons were extracted from a circular region of $\sim 46
\arcsec$ in radius. We have selected single, double, triple and
quadruple events. In the case of the MOS data, significant
photon pile-up was evident so the central $12\arcsec$ at the core of
the PSF were discarded from our analysis. The remaining count rates in
the 0.2--10 keV band are 5.6 and 5.5 c/s for MOS1 and MOS2
respectively.

\begin{figure}
\psfig{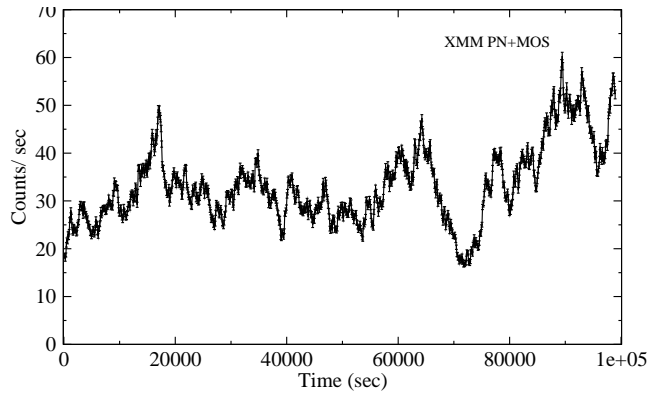}
\caption{\xmm\ combined PN and MOS, 0.2--10 keV light curve, binned at 96 s.}
\label{xmm_lc}
\end{figure}

To construct the light curves, we combined the data from the 3 EPIC
detectors and selected energy ranges to match the average energies of
the \asca\ light curves. The 0.7--2, 2--5, 5--10 and 2--10 keV \asca\
energy bands have the same mean energy as the 0.9--2, 2--4.5, 5--8 and
2--5.7 keV energy bands of the PN camera, respectively. We used these
PN energy bands for all EPIC detectors, as the PN counts dominate over
those of both MOS cameras. The combined 0.2--10 keV light curve,
binned to 96 s resolution, is shown in Fig.  \ref{xmm_lc}.

\subsection{\asca}

We used data taken by \asca\ during its long observation of \ark ,
between 2000 June 1 and July 5. The data was reduced as detailed in
\citet{Papadakis_ark} and we constructed light curves in the 0.7--2
(soft), 2--5 (medium) and 5--10 keV (hard) energy bands for all four
detectors, SIS0, SIS1, GIS2 and GIS3. As these data contain regular
gaps, due to the Earth occultation of the satellite, we binned the
data in orbit-long bins ($\sim 5400$ s) to obtain an evenly sampled
light curve, containing 551 points. To check the stability of the
detector through this month-long observation, we compared the ratios
between the light curves from different detectors. While SIS0, GIS2
and GIS3 showed consistent light curves, we observed discrepancies
between these and SIS1 in all energy bands. The ratio between the SIS1
light curve and the light curves from all other detectors shows a
linearly decreasing trend, of amplitude $\sim10$\% as measured from
the start to the end of the observation. We therefore combined only
SIS0, GIS2 and GIS3 data, in each energy band, to produce the final
light curves.

\begin{figure}
\psfig{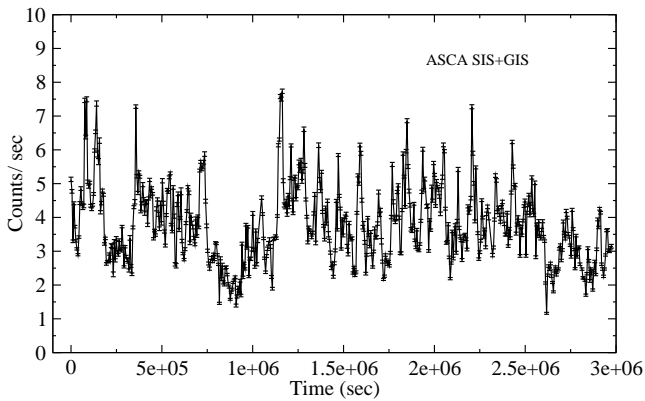}
\caption{\asca\ light curve of \ark\ in the 0.7--10 keV band, combined data from SIS0, GIS2 and GIS3. The light curve has been binned at 5408 s, which is approximately the orbital period of the satellite.}
\label{asca_lc}
\end{figure}

The combined, binned and background-subtracted $0.7--10$ keV light
curve is shown in Fig. \ref{asca_lc}. The average count rates for the
soft, medium and hard light curves are 2.7, 0.82 and 0.24 counts/s
respectively and the average exposure fraction is 23\%. The PSD and
other variability properties of this data set have been studied by
e.g. \citet{Papadakis_ark,Edelson}.

\section{Broad band coherence and lags}
\label{xs}

Fluxes in different energy bands may vary in a similar way and can do
it simultaneously, or with a delay. { This relative behaviour can be
studied by cross correlating two light curves, observed simultaneously
in different energy bands. The cross correlation measures the the
degree of linear correlation between the energy bands: if this
function presents a significant peak (reaching a value of 1) then the
light curves are well correlated and average delay between
fluctuations in both light curves is given by the position of this
peak on the lag axis. The degree of correlation and the length of the
lag might be different for different variability time-scales,
therefore
 it is more convenient to calculate the cross spectrum, which
calculates the coherence and time lag for each separate
time-scale. This technique uses the Fourier transform of the light
curves to separate the different Fourier components at the individual
frequencies, and measures the relative phases between the Fourier
components of the different light curves. If the relative phases at a
given Fourier frequency remain constant between different time
segments of the light curves, then the light curves are coherent at
this frequency, producing a coherence value of 1. On the other hand,
if the phases vary randomly between time segments, the coherence will
tend to 0. If the light curves are coherent, their relative phase,
i.e. phase lag, represents the delay between similar fluctuations in
both energy bands, which can be directly converted into a time lag see
section 3 of \citep[see section 3 of][for an in depth discussion on
the meaning of coherence and time lag functions]{Nowak_lags} .}

Time lags and coherence between two simultaneous time series
$s(t),h(t)$ can be estimated using the cross spectrum $C(f)=
S^*(f)H(f)$, where $S(f)$ and $H(f)$ are the Fourier transforms of the
respective light curves. The coherence $\gamma ^2$ for discretely
sampled time series is calculated as follows:

\begin{equation}
\gamma ^2 (f_{i})=\frac{ \langle {\rm Re}_C (f_{i}) \rangle ^2 +
\langle {\rm Im}_C (f_{i}) \rangle ^2}{\langle | {\rm S}(f_{i})|^2
\rangle \langle | {\rm H}(f_{i})|^2\rangle } ,
\end{equation} 
where ${\rm Re}_C (f_{i})$ and ${\rm Im}_C (f_{i})$ are the real and
imaginary parts of the cross spectrum $C(f)$ and angle brackets
represent averaging over independent measurements, either at
consecutive frequencies in a frequency bin or equal frequencies from
different light curve segments.

The argument of the cross spectrum defines the phase lags:
$\phi( f_{i})=\arg{\langle C(f_{i})\rangle}$, and from here the time lags,
$\tau(f_i)$, are calculated as:

\begin{equation}
\tau (f_{i})= \frac{\phi(f_{i})}{2\pi f_{i}}= \frac{1}{2\pi f_{i}} 
\arctan \left\{ \frac{ \langle {\rm Im}_C (f_{i}) \rangle }
{\langle {\rm Re}_C (f_{i})\rangle } \right\}
\end{equation} 

The cross spectrum produces estimates of the coherence and time lags
as a function of Fourier frequency, or equivalently, of time-scale.
\citet{Vaughan_coh} and \citet{Nowak_lags} discuss the interpretation
of these measurements in detail and provide error estimates for data
with observational noise. We used the methods described therein to
estimate the error bars on the time lags.

\subsection{Coherence}

We used \asca\ data to compute lags and coherence in the $10^{-6}
-10^{-4}$ Hz frequency range and \xmm\ data for the $10^{-4} -10^{-2}$
Hz frequency range. { Although the data sets were taken 5 years
  apart, they can be combined as AGN are not expected to change
  `state' on time-scales of less than $\sim $ 1000 years, assuming
  linear scaling of the time-scales seen in BHXRBs by the respective
  black hole masses. This assumption is supported by the fact that the
  PSD does not change between the observations, at least in the
  frequency region where the observations overlap (McHardy et al. {\sl
    in prep.})  and, as will be shown, the cross spectrum properties
  also show continuity from one data set to the other.}

As a first step we used the $0.7-2.0$ keV and $2.0-10.0$ keV \asca\
bands to obtain the highest possible signal-to-noise in the light curves. As
the lags and coherence are often energy-dependent, we used \xmm\ light
curves with the same
\emph{average energies} as the \asca\ bands (see Section~2.1). 

\begin{figure}
\psfig{figure=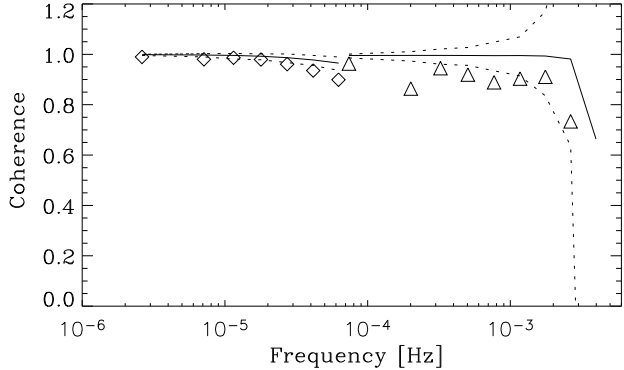,width=8.5cm,height=5cm}
\caption{Coherence between 0.7--2 and 2--10 keV \asca\ energy bands and matching \xmm\ bands, as a function of Fourier frequency. \asca\ and \xmm\ lag estimates are plotted in diamonds and triangles, respectively. The solid line represents the median of the distribution of coherence values from simulated data, as described in Sec. \ref{significance}, the dotted lines represent the top and bottom 95\% extremes of this distribution. The simulated light curves have intrinsic coherence of unity. The measured coherence follows the behaviour of the median of the distribution of simulated data, indicating that the drops at the highest frequencies for each data set are (partly) due to noise limitations.}
\label{coh_test}
\end{figure}

{ We calculated the cross spectrum for each pair of light curves
and binned the coherence and phase lag estimates in logarithmically
spaced frequency bins, including a minimum of 10 points per bin.}  The
resulting coherence function is shown by the markers in
Fig.~\ref{coh_test}, where diamonds denote \asca\ data points and
triangles denote \xmm\ points. The coherence measurements in this plot
have been corrected for Poisson noise effects in the high
signal-to-noise limit given by \citet{Vaughan_coh}.  The solid and
dotted lines represent the scatter expected for each data set,
estimated under the assumptions that will be explained in
Sec.~\ref{significance}. The measured coherence is high ($\sim 0.9$)
for the entire frequency range up to $\sim 10^{-3}$Hz. At higher
frequencies, the coherence drops drastically (the highest-frequency
point is negative, not shown in the plot) but the expected scatter
increases significantly, making coherence measurements in this range
unreliable.

\subsection{Time lags}

\begin{figure}
\psfig{figure=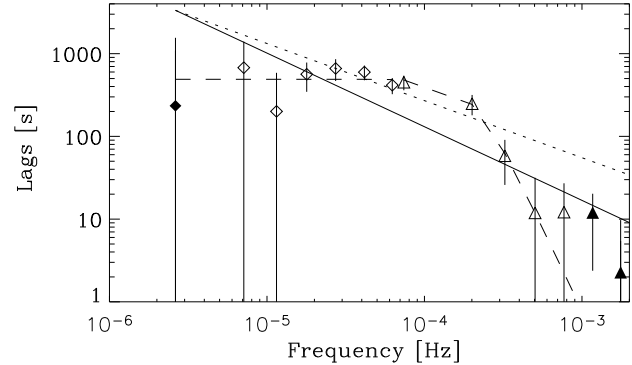,width=8.5cm}
 \caption{Lag spectrum of \ark\ calculated using \asca\ data (diamonds) and \xmm\ data (triangles) between the 0.7--2 and 2--10 keV \asca\ energy bands, plotted as a function of Fourier frequency. The best-fitting power law model for the entire frequency range plotted here ($\tau(f)=0.04f^{-0.9}$) is shown by the solid line. The dotted line represents the best-fitting power law fit to the $10^{-5} -5\times 10^{-4}$ Hz frequency range ($\tau(f)=0.5f^{-0.7}$) and the dashed line represents the best fitting bending power law model, with bend frequency of $2\times 10^{-4}$ Hz and low and high frequency slopes of 0 and -4 respectively. The error bars were calculated following the procedure described by \citet{Nowak_lags}.}
\label{lags_tot}
\end{figure}

 Figure \ref{lags_tot} shows the lag spectrum over the frequency
 range
 where the measured coherence is high, i.e. below $\sim 2
 \times
 10^{-3}$ Hz, as lags measured in cases of low coherence are
 not
 meaningful.  Positive lag values indicate that the soft band
 leads the
 hard. Significant lags are detected between $\sim 10^{-5}
 -10^{-3}$
 Hz. The \xmm\ and \asca\ lag spectra match well at the
 frequencies
 where they overlap, around $10^{-4}$ Hz. The lag
 spectrum appears to
 be frequency-dependent, where larger time lags
 are associated with
 longer time-scale fluctuations, similar to what
 has been observed in
 other AGN and BHXRBs (see
 Sec.~\ref{comparison}). The lags spectrum in
 the frequency range
 between $2\times 10^{-5}$ and $2 \times 10^{-4}$ Hz resembles the
 shape of a power law. However, at lower and higher frequencies, the
 measured lags decrease noticeably below the extension of a power law
 fitted to the lag spectrum in this central frequency range.

A single power law fit to the lag spectrum over the whole frequency
range shown in Fig.~\ref{lags_tot}, yields a best-fitting model
$\tau(f)=0.04f^{-0.9}$, with a $\chi^2=62$ for 14 dof (solid line in
Fig.~\ref{lags_tot}). This is clearly an unacceptable fit and leaves
the best-defined lag measurements well above the fitted
curve. Restricting the fitting range to $\sim 10^{-5}-5\times
10^{-4}$
 Hz produces a similar power law, $\tau(f)=0.5f^{-0.7}$, with
a
 $\chi^2=9.9$ for 5 dof, shown by the dotted line in the same
figure. The measured lags above and below this range fall far below
the extension of the power law fit, so the lag spectrum resembles a
broad hump. A much better fit to the entire frequency range was
obtained by using a single-bend power law model. { We chose this
model to replicate a power law that bends gently from one slope,
$\alpha_L$, to another, $\alpha_H$, around a bend frequency $f_b$,
given by:}
\begin{equation}
\label{psd_eq}
\tau(f)=A f^{\alpha_L}/(1+(f/f_b)^{\alpha_L-\alpha_H}).
\end{equation}
The best fitting values for the low frequency and high frequency
slopes ($\alpha_L$ and $\alpha_H$ respectively) are 0 and -4, while
the bend is at a frequency of $2 \times 10^{-4}$ Hz. This flat time lag
spectrum, bending to a very steep high frequency slope model produces
a $\chi^2= 10.9$ for 16 dof.

The significance of the deviations from a simple power law and the
goodness of fit of the bending power law model were assessed through
the Monte Carlo simulations discussed in the following section.

\subsection{Estimation of the significance}
\label{significance}

A power law lag spectrum could be distorted by sampling and
observational noise effects. We used Monte Carlo simulations to test
the possibilities that the measured lag spectrum deviates from a power
law only by these effects and that the drops observed in the coherence
function are caused by observational limitations on intrinsically
coherent light curves.

We constructed 2000 light curve pairs to simulate simultaneous hard
and soft light curves, using the method of \cite{Timmer}. We used a
double-bending power law model for the underlying PSD, with the parameters
found by McHardy et al. ({\sl in prep.}):
\small
\begin{equation}
P(f)=Af^{-\alpha_L}\left[1+\left(\frac{f}{f_{bL}}\right)^{\alpha_M-\alpha_L}\right]^{-1}\left[1+\left(\frac{f}{f_{bH}}\right)^{\alpha_H-\alpha_M}\right]^{-1} 
\end{equation}
\normalsize where $A=3\times 10^{4}, \alpha_L=0, \alpha_M=1.2,
\alpha_H=4.5, f_{bL}=8.7\times 10^{-7} $ Hz and $f_{bH}=2 \times
10^{-3}$ Hz. By construction, the light curve pairs have a coherence
of unity at all frequencies. Time-scale dependent lags were introduce
by shifting the
 phase component of the Fourier transform of the
`hard' simulated light
 curves, by $2\pi f \tau(f)$. Appropriate
Poisson noise was added to
 the resulting simulated light curves.

\asca\ data simulations were generated in 100~s bins and sampled
in
 exactly the same way as the real 100~s binned light curve. They
were
 subsequently re-binned in 5400~s evenly spaced bins, just as
was
 done for the real data. Unlike \asca\ data, the real \xmm\
light
 curves are practically continuously sampled. Therefore,
\xmm\
 simulated light curves were simply generated in 24 s bins and
then
 re-binned in 96 s bins. 

\begin{figure}
\psfig{figure=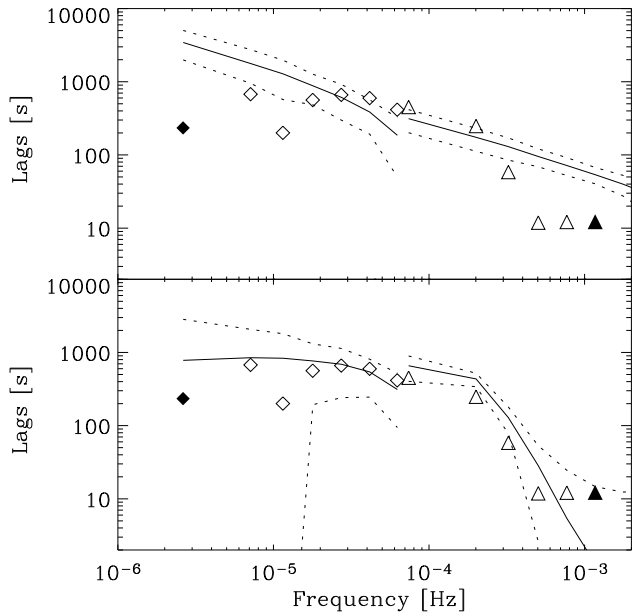,width=8.5cm,height=9cm}
\caption{Top: Distribution of simulated lag spectra, using $\tau(f)=0.5f^{-0.7}$ as the underlying lag model, as described in Sec. \ref{significance}. The solid line represents the median and the dotted lines the 95\% extremes of the distribution of simulated data. Lags of \ark\ calculated between 0.7--2 and 2--10 keV \asca\ energy bands and matching \xmm\ bands are plotted in diamonds (\asca\ data) and triangles (\xmm\ data). Filled symbols denote \emph{negative} lags of the amplitude shown. Below $2 \times 10 ^{-5}$ Hz and above $\sim 3\times 10^{-4}$ Hz, the measured lags fall below the the 5\% lowest simulated lag values, indicating that there is a real break in the lag spectrum. Bottom: Same as above but using a bending power law model as the underlying lag spectrum.}
\label{lags_test}
\end{figure}

The cross spectrum for each pair of simulated light curves was
computed using the same binning used for the real data. The median of
the distributions of coherence and lag values of the simulations, for
each Fourier frequency, are plotted in solid lines in
Figs. \ref{coh_test} and \ref{lags_test}. The dotted lines in the
same
 figures mark the spread of the distribution of simulated values
so
 that 5\% of the points lie above the top line and 5\% lie below
the
 bottom line.

The measured coherence in Fig. \ref{coh_test} follows the trend of
the
 median of the distribution of simulations, for both data
sets. The
 small coherence drop on the highest \asca\ frequency bins
is probably
 partly due to inaccurate Poisson noise corrections, as
suggested by
 the simulations (the formula for this correction
derived by
 \citet{Vaughan_coh} is not strictly applicable when the
variability
 signal-to-noise is low). Above this frequency, the
coherence drops
 slightly below the distribution of simulated data,
indicating a real
 but small ($<$10\%) loss of coherence. Finally,
our results suggest
 that the strong drop above $10^{-3}$ Hz can be
easily explained by
 Poisson noise effects, and hence is most
probably not intrinsic.

As for the phase shift in the Fourier
components of the two light
 curves, initially we used
$\tau(f)=0.04f^{-0.9}$, i.e. the best fit to
 the lag spectrum over
the entire frequency range, as the underlying
 lag spectrum.  As seen
in Fig. \ref{lags_tot} this assumed lag
 spectrum falls well below
the best-defined lag measurements in the
 middle of the frequency
range probed. Not surprisingly, the four
 central data points remain
above the top 95\% of the distribution of
 simulated lags, while two
high-frequency points still fall below it,
 implying that the
underlying lag spectrum is inconsistent with the
 simple power law
fitted to the data. We repeated the test using
 $\tau(f)=0.5f^{-0.7}$
as the underlying lag spectrum, i.e. the fit to
 the $\sim
10^{-5}-5\times 10^{-4}$ Hz frequency range. The resulting
 lag
spectra distribution is plotted in Fig. \ref{lags_test}.  The drop
in the lags at low and high frequencies is significant as the lags at
the extreme frequencies fall far below
 the 95\% lower limit of the
distribution of simulated data.  We note,
 also, that the artificial
high-frequency break in the lag spectrum,
 produced by sampling
effects and noticed by \citet{Crary} cannot
 explain the break we
observe. The artificial break should appear
 around $0.5\times f_{\rm
N}$ where $f_{\rm N}$ is the Nyquist
 frequency while, for the \xmm\
data we used, $0.5 \times f_{\rm N}
 \sim 2.6\times 10^{-3}$ Hz, and
the break we observe in \ark\ is at an
 order of magnitude lower
frequency, at $\sim 2\times 10^{-4}$ Hz.

 At the low frequency end, the median of the distribution of
 simulated
 lag spectra does not reproduce the decreasing trend seen
 in the data
 and many data points fall below the 95\% lower
 limit. The small
 scatter expected in the lag measurements is partly
 due to the large
 lags intrinsic to the underlying model we
 assumed. We repeated the
 same test, this time using the bending
 power law as the underlying lag
 spectrum. The small lags at low
 frequencies, that this model produces,
 increases the low frequency
 scatter significantly. Thereore, a single bend at
 high frequencies
 can account for both bends, including the
 negative lag values of the
 data at the lowest frequencies. Therefore,
 a single-bend model, with
 constant time lags below $\sim 10^{-4} $ Hz
 is consistent with the
 data. The median and 95\% extremes of the
 distribution of lag values
 for the bending power law model are shown
 in the bottom panel of
 Fig. \ref{lags_test}.

For completeness, we
calculated lag spectra using data from the long
 monitoring campaign
of \ark\ performed with \rxte, described by
\citet[e.g.][]{Pounds}. The lag spectra obtained from these \rxte\
data are inconclusive however, as the low variability power below
$\sim 10^{-6}$ Hz produces a very weak signal in this frequency
range,
 making lag measurements too uncertain.
 
 We conclude that
the changes in slope of the lag spectrum at $\sim
 2\times 10^{-5} $
and $\sim 2\times 10^{-4} $ Hz are significant, so
 the spectrum is
not consistent with a single power law model. The
 spectrum is
consistent with a bending power law but we cannot assess
 accurately
the behaviour of the lags below the bend frequency. A model
 of
constant lag up to $10^{-4}$ Hz bending to a $f^{-4}$ dependence at
high frequencies can reproduce the data well.

\section{Energy dependence of the lags}
\label{energy} 
\begin{figure}
\psfig{figure=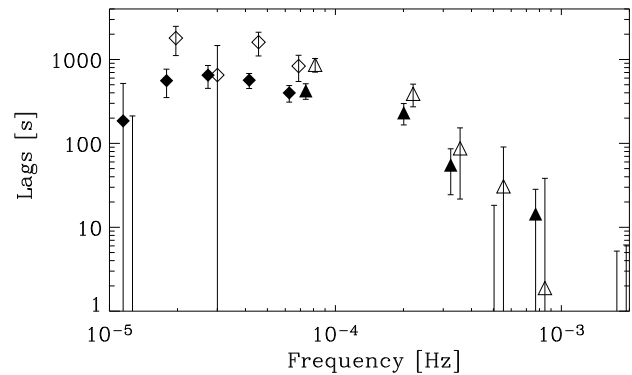,width=8.5cm,height=5cm}
\caption {Time lags between 0.7--2 keV  and 2--5 keV (filled diamonds) and 5--10 keV (open diamonds) \asca\ light curves. Lags between the corresponding EPIC bands are plotted in filled and open triangles. The lag spectra retain their shape but the amplitude of the lags increase by a factor of $\sim 2$ for a factor of 3 increase in the energy separation of the bands. For clarity, open symbols have been shifted to slightly higher frequencies, their true frequencies being identical to the ones shown by the filled symbols. }
\label{lags_bands}
\end{figure}

The magnitude of the time lags between energy bands tends to increase
with energy separation, in both AGN and BHXRB
\citep[e.g.][]{Papadakis_7469, Nowak_lags}. In \ark , this effect is
clearly observable over the frequency range where significant lags can
be measured. Figure \ref{lags_bands} shows the lags between soft and
medium and soft and hard \asca\ bands, together with the corresponding
\xmm\ bands. An increase of the energy separation in the bands used
to make the cross-spectrum, from a factor 3 to a 
factor of 6 difference in average
energies, produces an increase in the lags by a factor of $\sim 2$,
while preserving the shape of the lag spectra, within the
uncertainties.

\begin{figure}
\includegraphics[width=8.5cm]{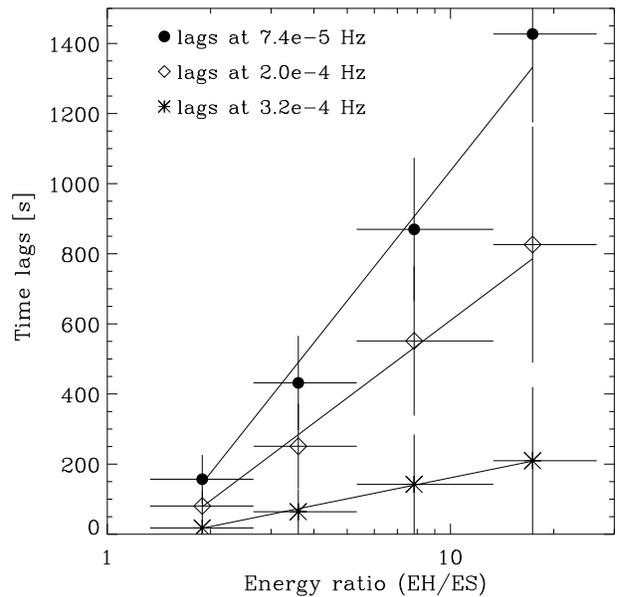}
\caption{Time lags estimated using the \xmm\ light curves, between the 0.2--0.5 keV band and the 0.5--1, 1--2, 2--5 and 5--10 keV hard bands, at three temporal frequencies, $7.4\times 10^{-5}$ , $2\times 10^{-4}$ and $3.2 \times 10^{-4}$ Hz. The markers are set at the average-energy ratios of the photons in each band, the horizontal lines mark the width of each energy band ratio and are not error bars. The error bars on the lag values were calculated with the prescription of \citet{Nowak_lags}, notice that this might be an overestimate of the \emph{relative error} between different energy bands of simultaneous light curves. The lags increase linearly with the logarithm of the energy separation.}
\label{lags_vs_e}
\end{figure}

To investigate the energy dependence in more detail, we used the
broader energy bandpass of \xmm\ to construct light curves in 5
different bands, 0.2--0.5, 0.5--1, 1--2, 2--5 and 5--10 keV. The
time
 lags between the 0.2--0.5 keV soft light curve and all the
harder
 light curves are shown in Fig. \ref{lags_vs_e}. Here the lags
are
 plotted as a function of the average-energy ratio between each
hard
 band and the 0.2--0.5 band ($E_H/E_S$), for the three lowest
Fourier
 frequency bins, which had the smallest error bars. In all
cases, the
 lags increase linearly with the logarithm of the
energy. The
 best-fitting relations, plotted in solid lines in
Fig. \ref{lags_vs_e}, have functional forms: $\tau(E)=1233 \log
E_H/E_S -196$ for $7.4\times10^{-5}$ Hz, $\tau(E)= 735 \log E_H/E_S
-
 125$ for $2.0\times10^{-4}$ Hz and $\tau(E)= 199 \log E_H/E_S -38$
for
 $3.2\times10^{-4}$ Hz. The errors in the lag values take into
account the red-noise nature of the light curves and probably
overestimates the \emph{relative} error between different energy
bands. Therefore, the $\chi^2$ values for the fits are not significant.
A perfect log-linear relation between lag
 amplitude and energy ratio
should cross (1,0) in this plot, as the
 lags should tend to 0 when
calculated between identical energy
 bands. Therefore, the log-linear
relation might not hold down to small
 energy differences. This
possible change in the energy dependence is
 not significant however,
given that if we force the intercepts to
 equal 0, the fits are still
acceptable.

A similar linear dependence of the lags on the logarithm of the energy
ratio has been observed for the BHXRB Cyg~X-1 \citep{Cui,
  Crary,Nowak_lags}, suggesting that a similar mechanism operates in
both sources to produce the time lags.

\section{Discussion}
\label{discussion}

{ The soft (0.7-2 keV) and hard (2-10 keV) X-ray light curves measured for \ark\ show highly coherent variability, at least over time-scales of $10^{-6}-10^{-3}$ Hz. These energy bands are dominated by two distinct energy-spectral components, a soft excess dominating up to $\sim 2$ keV and a hard power law dominating at higher frequencies \citep{Turner}. These authors have shown, by time-resolved spectral fittings, that these components vary in a similar fashion, albeit with different amplitudes, down to time-scales of at least a day. Given that both spectral components vary, our results of high coherence confirm that the soft excess and the power law variability are well correlated and show that the correlation holds down to time-scales of 1000 s.}

\subsection{Comparison with other AGN lag spectra}
\label{comparison}

Significant time lags between X-ray energy bands have been measured
for a few AGN, NGC~7469 \citep{Papadakis_7469}, MCG--6-30-15
\citep{VaughanMCG}, NGC~4051 \citep{McHardy4051} and NGC~3783
\citep{Markowitz3783}. In all cases, the lags appear to increase with
variability time-scale and, where it has been possible to measure,
also with energy separation of the bands.  Previous lag spectra were
normally fitted with single power law models and, in many cases, only
the amplitude of the lags could be left as a free parameter, due to
the quality of the data. Consequently, even if there were intrinsic
slope changes in the lag spectra of any other AGN, they would not have
been detected.

To allow a direct comparison of the amplitude of the lags in different
objects, we calculated lag spectra for NGC4051, MCG--6-30-15 and
NGC3783 using archival \xmm\ data in the same PN energy bands as used
for \ark , i.e. 0.9--2 and 2-4.5 keV. We used these energy bands to
plot our \ark\ \xmm\ lag spectra together with the \asca\ data,
matching the average energies of the bands. These AGN lag spectra are
shown in the top panel of Fig. \ref{lag_spectra}.  The plotted lags
are smaller than those published for these objects by other authors
because we used closer energy bands and the lags tend to increase with
the separation of the energy bands. We plotted fractional lags, given
by time-lag/time-scale (equal to $\phi(f)/2\pi$), versus frequency in
terms of the PSD break frequency for each object. The PSD break
frequency estimates that we used are listed in Table \ref{agn_fbs} and
were taken from literature (references are listed in the same
Table). Notice that a constant \emph{time} lag translates into
increasing \emph{phase} or \emph{fractional} lags.

The lags in \ark , plotted in filled circles in
Fig. \ref{lag_spectra}, reach noticeably larger values than all the
other objects. Compared to NGC4051 (open diamonds), the shapes and
amplitudes are similar in the region where they overlap. At higher
relative frequencies, MCG--6-30-15 (filled triangles) and NGC3783
(stars), show much smaller lags. These two objects do show noticeable
lags for larger separation of the energy bands, but for the same bands
used for \ark\ their lags are too small to be detected clearly and
their spectra are consistent with 0 within the errors. It is possible
that in all the objects plotted, the size of the lags drops towards
the break frequency in the PSD. If this is the case, the large lags
detected in \ark\ effectively probe a different part of the lag
spectrum and we would need lag measurements at lower frequencies in
the other AGN to make a significant comparison.

\subsection{Comparison with BHXRBs}

\begin{figure}
\psfig{figure=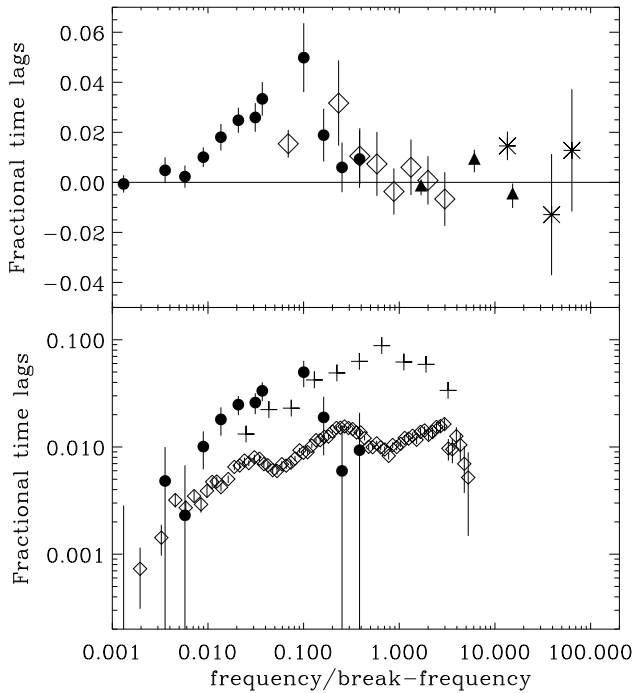,width=8.5cm,height=9.5cm}
\caption{Top panel: Fractional lag (time lag/time-scale) spectrum for
\ark , in filled circles, compared to lag spectra in the same energy
bands of NGC4051 (open diamonds), MCG--6-30-15 (filled triangles) and
NGC3783 (stars). The fractional lags are plotted as a function
frequency normalised to the PSD high frequency break for each object,
as explained in Sec. \ref{comparison}. A linear scale was used for the
y-axis to allow negative lag points to be plotted. Bottom panel: same
as above but comparing with the BHXRB Cyg~X-1 in the hard state
(diamonds) and one intermediate state (crosses), the PSD break
frequencies used were 6 Hz and 10 Hz, respectively.}
\label{lag_spectra}
\end{figure}

Figure \ref{lag_spectra} also shows a comparison of the \ark\ lag
spectrum with the lag spectra of Cyg~X-1 in the low/hard and
intermediate states.  In the low/hard state, plotted in diamonds,
Cyg~X-1 shows significant fractional lags over
 more than two decades
in frequency. These data correspond to an {\it RXTE} long-look
observation taken on 1996 Dec 16--19 (total useful exposure
 time
$\sim89$~ks) and we used $2^{-8}$~s resolution light curves in
 soft
(2--4 keV) and hard (8--13~keV) energy bands. We measured a PSD break
frequency at 6 Hz and used this value to scale the lag spectrum shown
in Fig.~\ref{lag_spectra}.  A step-like structure is clearly visible
in the low/hard state lag spectrum, which, as noted  by
\citet{Nowak00}, may be explained if each step corresponds to the lag
value of individual broad Lorentzian components in the PSD.  A lag
spectrum of Cyg~X-1 in a transition state is plotted in crosses in
the bottom panel of Fig. \ref{lag_spectra}. For this plot we used
\rxte\ data taken on 2000 Nov 3 (ObsID 50119-01-03-01) in the 2--5.7
and 9.4--15 keV energy bands, measuring the PSD break at a
frequency of 10 Hz.  In this case, the lag spectrum is more clearly
`peaked' than in the low/hard state, and the lags are significantly
larger (e.g. as noted by \citealt{Pottschmidt}).  Lag spectra of a
similar peaked shape were found by \citet{Miyamoto} in the BHXRB
GX~339-4 in the very high state and in the X-ray nova GS1124-68 in
the `very high flare state' defined in the same paper.

Although the peaked shape of the \ark\ lag spectrum resembles slightly
one of the steps seen in Cyg~X-1 low/hard state, the drops in the case
of \ark\ are much more pronounced and the fractional lag spectrum peak
is correspondingly narrower.  As \citet{Pottschmidt} have shown, the
lags in the high/soft and low/hard states of Cyg~X-1 are very similar
in magnitude and broad time-scale dependence, so the \ark\ lag
spectrum does not correspond to that of a high/soft state either.  The
strong drops in the lag spectrum of \ark , together with the larger
amplitude of the lags, show that the data resemble the lag spectrum of
Cyg~X-1 in the intermediate state (or equivalently, the very high
state in BHXRB transients, where similar properties are seen at higher
luminosities).  Therefore, the spectral-timing data supports the
earlier suggestion by \citet{Papadakis_ark} that Ark~564 is similar to
a BHXRB in the {\it very high state}.  Although the peak frequencies
of the \ark\ and Cyg~X-1 lag spectra are different when scaling by the
PSD break frequencies, this difference may reflect the fact that the
break-frequencies defined for Cyg~X-1 are based on Lorentzian fits to
the data, and not the highest-frequency cut-off observed in the PSD,
which is probably not detectable due to the low signal to noise at
high frequencies in the PSD.

\begin{table}
\small
\begin{tabular}{ccl}
\hline
Object& PSD break & reference\\
&frequency [Hz]\\
\hline
NGC 3783& $4\times10^{-6}$& Markowitz et al. 2003\\
NGC 4051&$8\times10^{-4}$ &McHardy et al. 2004\\
MCG--6-30-15&$6\times10^{-5}$&McHardy et al. 2005\\
\ark\ &$2\times10^{-3}$& McHardy et al. {\sl in prep.}\\
\hline                   
\end{tabular}
\caption{\label{agn_fbs} PSD break frequencies used for Fig. \ref{lag_spectra}}
\end{table}

\subsection{Interpretation in terms of variability models}

The lag spectrum of \ark\ shows a constant absolute time-lag over more
than a decade range in Fourier frequency, which cuts off at high
frequencies, equivalent to a peaked shape in the fractional time-lag
spectrum.  This shape suggests that only a single variability
component, with a single time-lag dominates the variability up until
the lag cut-off frequency.  It is interesting that the lag
cut-off does not correspond to the high-frequency turnover in the PSD,
but occurs a decade lower in frequency.  This result in turn suggests
that a second high-frequency component, with a much smaller (or zero)
lag dominates above the lag turnover, and produces the variability
observed between $2\times10^{-4}$~Hz and $2\times10^{-3}$~Hz.  We now
consider how this model for the lags can be physically realised.

{ The X-ray energy spectrum of \ark\ is composed of a soft excess
which dominates below 2 keV and a power law, which dominates at higher
energies \citep{Turner}. Time lags between soft and hard X-rays could
be produced if the soft excess variability leads that of the power
law. The energy dependence of the lags, however, indicates that the
soft excess cannot lead the power law component as a whole. As shown
in Fig. \ref{lags_vs_e}, the length of the lags increases with energy
separation, so that even energy bands that are fully dominated by the
power law component do show different lags. Therefore, the lags must
arise at least partly within the power law spectral component.}

Several schemes have been proposed to explain the origin of X-ray time
lags observed in BHXRBs and AGN. Notably, lags through Comptonisation
of seed photons into X-rays of different energies through different
numbers of scatterings can produce the observed logarithmic dependence
of the lags on energy ratio \citep[see for example discussion in][
section 5.3]{Nowak_lags}. { In the case of \ark\ the soft excess
photons could serve as input seed photons for a Comptonising corona
that would then re-emit them as a power law component, explaining the
lag between both components and the increasing lag with energy
separation within the power law itself.} This simple scheme is
challenged by timing considerations, however, given that the larger
number of scatterings that the high energy photons must go through
reduces their high frequency variability, compared to that of lower
energy photons, contrary to what is observed \citep[as noted by
e.g.][]{VaughanMCG}. { This scenario is also challenged by the
frequency dependence of the lags since lags arising simply through
Comptonisation would have the same value for all variability
time-scales. Therefore, the time lags between the various energy bands
most probably arise, at least partly, from the nature of the physical
process or by the geometry of the source which is responsible for the
hard band power law component}. A Comptonisation origin of the lags
might still be possible if the Comptonising medium has a
radially-dependent electron density, as proposed by \citet{kazanas}.
Since the turnover in the \ark\ lag spectrum suggests two separate
variability processes with different lags, these might be produced in
separate emission regions with different Comptonising structures.

Another possibility is that the lags are produced by accretion rate
fluctuations travelling inward through the accretion flow.  In this
case, the variability fluctuations can be produced quite far out in
the accretion flow, so low-frequency variations can still be observed
even if most of the X-ray emission originates in the centre of the
accretion flow, close to the black hole \citep{Lyubarskii}.  If the
locally emitted spectrum of the accretion flow hardens towards the
centre then the fluctuations are seen first in lower energy bands and
later in higher energy bands, which produces the observed hard
lags. This scenario was proposed by \citet{Kotov} and further
discussed by several authors
\citep[e.g][]{VaughanMCG,McHardyMCG,arevalo} and appears to be
consistent with the observed spectral-timing properties of the X-ray
light curves, as well as their non-linear nature \citep{umv}.  In
particular, the propagating fluctuation model can simultaneously
reproduce the high observed coherence, as well as reproducing the lag
amplitude and general lag spectral shape and energy dependence of the
PSD which is observed in BHXRBs \citep{arevalo}. { The required
hardening of the energy spectrum towards the centre could be realised
by having the soft excess spectral component emitted by an extended
region and the power law component produced only close to the
centre. A more detailed analysis of the energy spectrum in different
variability time-scales would be needed to test this possibility.}

In the propagating fluctuation model, a single variability component
can dominate the lag spectrum over a broad frequency range if it is
produced in a single annulus in the accretion flow, so that the
propagation time-scale through the emitting region (and hence the lag)
is the same for all variability time-scales.  Thus the lag-spectrum of
\ark\ might be produced if the fluctuations arise in two regions in
the accretion flow, one, producing variations below $10^{-4}$~Hz at
large radii, with detectable lags, and one producing higher-frequency
variations at small radii with a negligible lag. If the fluctuation
time-scales correspond to the viscous time-scale in a geometrically
thick accretion flow, the radii of origin for the low and
high-frequency variability components are respectively, of tens of
gravitational radii and a few gravitational radii, for the expected
black hole mass \citep{Botte} of a few $10^{6}$~M$_{\odot}$.

\section{Conclusions}
\label{conclusions}

We used new \xmm\ data and combined it with a long \asca\ observation
to calculate the lags and coherence of \ark\ over the broadest range of
time-scales obtained so far for any AGN. The length of the \asca\
observation and the good time resolution of the \xmm\ data,
allowed us to produce an accurate estimate of the coherence and the
time lags between different X-ray energy bands.

The coherence was close to unity from $10^{-6}$ down to $\sim 10^{-4}$
Hz. Above this frequency, the coherence drops slightly below the
expected scatter inferred from simulations, to a value of $\sim
0.9$. The observed coherence drops by $\sim 10\%$ in the \asca\ data,
at $\sim 10^{-4}$ Hz and by $30\%$ above $\sim 10^{-3}$ Hz in
the \xmm\ data, i.e. at the shortest time-scales of each data set, are
at least partly due to observational biases.

We found significant time lags between different pairs of energy
bands, with harder bands lagging softer ones. The magnitude of these
hard lags increases with the energy separation of the bands, over the
entire frequency range tested. \xmm\ data was used to show that the
amplitude of the lags increases linearly with the logarithm of the
energy ratio between the soft and hard bands used, and that the
increase is stronger at lower temporal frequencies. The same
dependence has been observed before in BHXRBs
\citep[e.g][]{Nowak_lags}, implying that a similar mechanism operates
in both types of source to produce X-ray time lags.

The lag spectrum follows an approximate power law behaviour in the
frequency range $10^{-5}-5\times 10^{-4}$ Hz, with a best-fitting
slope of $\tau(f)=0.5f^{-0.7}$. Above and below this range, the time
lags drop to much lower values. We show that both drops are
significant, implying that the lag spectrum is inconsistent with a
single power law model, and there is, at least, one change in
slope. This broken lag spectrum resembles the lag spectra seen in
BHXRBs in the very high or transition state \citep[see
e.g.][]{Pottschmidt}, and is significantly different to the single
power-law lag spectra normally observed in other BHXRB
states. Compared with other AGN lag spectra, calculated between the
same energy bands, the lags in \ark\ are larger than all other
measurements. The shape and large amplitude of the lags spectrum match
well the description of BHXRBs VHS lag spectra as having strongly
enhanced lag values over a limited frequency range.

The band-limited lags observed in \ark\ can be interpreted in terms of
a few variability components, perhaps arising from localised annuli in
the accretion flow, with a single time lag associated to each of
them. If one of these components happens to dominate the variability
power over a broad range of time-scales, then its associated lag value
would extend throughout the same range. In the case of \ark , we note
that the two breaks seen in the PSD (\citealt{Pounds,Papadakis_ark},
McHardy et al. {\sl in prep.}) could correspond to mainly two broad
variability components contributing with a large fraction of the total
variability power. The lags (of $\sim 600$ s for the \asca\ energy
bands), seen up to $\sim 10^{-4}$ Hz would correspond to the PSD
component peaking at $10^{-6}$ Hz (low frequency PSD break) and a much
smaller lag value would be associated with the component peaking at
$10^{-3}$ Hz (at the high frequency PSD break). Therefore, the change
in slope seen in the lag spectrum can represent the drop from one lag
value to the other, or equivalently, the cross-over frequency of the
two main variability components.

\section*{Acknowledgements}

This work is based on observations with \xmm, an ESA science mission
with instruments and contributions directly funded by ESA Member
States and the USA (NASA). We wish to thank the anonymous referee for
helpful comments which improved the clarity of the paper. PA
acknowledges financial support from EARA network and the hospitality
of the Physics Department of the University of Crete and the Institute
of Astronomy, Cambridge. Part of this work was supported by the
bilateral Greek-German IKYDA2004 personnel exchange research project.

\bsp
\label{lastpage}
\end{document}